\documentclass{article}
\usepackage{graphicx}

\begin{document}
\date{}
\def\be{\begin{equation}}
\def\ee#1{\label{#1}\end{equation}}

\title{Irreversible processes in a Universe modelled as a 
mixture of a Chaplygin gas and radiation} 
\author{G. M. Kremer\thanks{\it Departamento de F\'\i sica, Universidade 
Federal do Paran\'a,
Caixa Postal 19044, 81531-990 Curitiba, Brazil}$\;$\thanks{E-mail: 
kremer@fisica.ufpr.br}}
\maketitle

\begin{abstract}
The evolution of a Universe modelled as a mixture of a Chaplygin gas 
and radiation is determined by taking into account
irreversible processes.
This mixture  could interpolate periods of a radiation dominated, a matter 
dominated and a cosmological constant dominated Universe. 
The results of a Universe modelled by this mixture are compared with the 
results of a mixture whose 
constituents are radiation and quintessence.
Among other results it is shown that: (a)  for both models there exists
a period of a past deceleration with a present acceleration; (b) the slope 
of the acceleration of the Universe modelled 
as a mixture of a Chaplygin gas with radiation is more pronounced than 
that modelled as a mixture of quintessence and 
radiation; (c) the energy density of the Chaplygin gas  tends to a 
constant value at earlier times than the energy 
density of quintessence does; (d) the energy density of  radiation for 
both mixtures 
coincide and decay more rapidly than the energy densities of the 
Chaplygin gas and of quintessence.

\end{abstract}

\noindent
Key words: Chaplygin gas, accelerated Universe, quintessence

Recent measurements of the anisotropy of the cosmic microwave background 
and of the  type 
Ia supernova SN 1997ff redshift indicate that the Universe is flat with a 
present positive acceleration and a past 
decelerating period~\cite{Per,Ri,TR}. Moreover, it has been recognised 
that a significant part of the energy density 
of the Universe is not due to matter or radiation but to an extraordinary 
non-baryonic matter and energy, 
which has a negative pressure and is the responsible for the present 
acceleration of the Universe. Two candidates for 
this energy, also called dark energy, are found in the literature. 
One of them is the quintessence with an equation
of state $p_X=w_X\rho_X$ and with the condition $w_X<-1/3$ (see, 
for example~\cite{Stein,Stein1,PR}), while the other refers to a 
Chaplygin gas with an exotic equation of state $p_c=-A/\rho_c$ 
subjected to the 
condition $A>0$ (see, for example~\cite{KMP,FGS,BBS,DAJ}). For 
some properties of the Chaplygin gas one is referred to the book
by Jackiw~\cite{JK}.

The irreversible processes in a homogeneous and isotropic Universe 
are normally described  by the so-called
viscous cosmological models which  are based on thermodynamic theories. 
There exist two kinds of thermodynamic theories: 
one within the framework of Eckart's (or first order)
thermodynamic theory (see, for example~\cite{Mur,Gro,KD,KD0,KD1}) where 
the non-equilibrium pressure is taken as a constitutive
quantity, while the other  considers an evolution equation for the 
non-equilibrium pressure within the framework of
the extended (or causal or second order) thermodynamic theory (see, 
for example~\cite{KD0,KD1,Be, RP,CJ,CH,Zim}). 

Recently~\cite{KD1} the evolution of a Universe modelled as a mixture 
of a matter field with quintessence was analysed 
by taking into account the irreversible transfer of energy densities 
of the matter and gravitational fields. 
Among other results, it was shown
that: (a) there exists a period of past decelerating followed by a  
present acceleration of the Universe 
due to the quintessence; (b) the
energy density of quintessence decays more slowly than the energy 
density of the matter field does.

The objective of this work is to analyse a Universe modelled as a 
mixture of a Chaplygin gas and radiation by taking
into account irreversible processes within the framework of extended 
thermodynamics. This mixture is more suitable than the one 
with radiation and quintessence as constituents, since - according 
to the work~\cite{KMP} - the Chaplygin gas can interpolate 
a matter dominated Universe (dust or pressure-less fluid) and a 
cosmological constant dominated Universe. Hence, the mixture we are 
interested in could interpolate a period of a radiation dominated, a 
matter dominated and a cosmological constant dominated Universe. 
This interpolation between the three periods is not possible for a 
Universe modelled by a mixture of radiation and quintessence.
The only possibility is to adjust in each period the barotropic 
equation of state of the matter field.

In this work we compare the 
results of a Universe modelled by a mixture of a Chaplygin gas and 
radiation with the results of a mixture
 whose constituents are radiation and quintessence.
Among other results it is shown that for both models there exists
a period of a past deceleration with a present acceleration while 
the slope of the acceleration of the Universe modelled 
as a mixture of a Chaplygin gas with radiation is more pronounced 
than that modelled as a mixture of quintessence and 
radiation. Moreover, the energy density of the Chaplygin gas  
tends to a constant value at earlier times than the energy 
density of quintessence does and  the energy density of  radiation 
for both mixtures 
coincide and decay more rapidly than the energy densities of the 
Chaplygin gas and of quintessence.

We consider the Robertson-Walker metric and model a spatially flat, 
homogeneous and isotropic
Universe as a mixture of a radiation field and an exotic fluid 
characterised by the so-called 
Chaplygin gas. The pressure of the radiation field $p_r$ and the 
pressure of the Chaplygin gas $p_c$ are related, respectively, 
to  their   energy densities $\rho_r$ and $\rho_c$ by
\be
p_r={1\over 3}\rho_r,\qquad p_c=-{A\over \rho_c},\qquad\hbox{with}
\qquad A=\hbox{constant}>0.
\ee{1}

For this kind of Universe, the energy-momentum tensor $T^{\mu\nu}$ 
of the sources that appear in Einstein's field equations is 
given by
\be
T^{\mu\nu}=(\rho_r+\rho_c+p_r+p_c+\varpi)U^\mu U^\nu-
(p_r+p_c+\varpi)g^{\mu\nu},
\ee{2}
where $U^\mu$ (such that $U^\mu U_\mu=1$) is the four-velocity  
and  $\varpi$ denotes the non-equilibrium pressure which  is 
responsible  for the dissipative effects during the 
evolution of the Universe.

The balance equation for the energy density of the mixture follows 
from  the conservation law of the energy-momentum tensor
${T^{\mu\nu}}_{;\nu}=0$ which in a comoving frame reads
\be
\dot\rho_r+\dot\rho_c+3H(\rho_r+\rho_c+p_r+p_c+\varpi)=0.
\ee{3}
Above, the over-dot  refers to a differentiation with respect to time 
and $H=\dot a/a$ is the Hubble parameter with 
$a(t)$ denoting the cosmic scale factor.

The Friedmann  equation  connects the evolution of the cosmic scale factor 
with the energy densities of the radiation field and of the Chaplygin 
gas, i.e.,
\be
H^2={8\pi G\over 3}(\rho_r+\rho_c),
\ee{4}
where $G$ is the gravitational constant.

We assume that the Chaplygin gas interacts only with itself and is 
minimally coupled to the gravitational field.
In this case, the  balance equation for the energy density of the 
Chaplygin gas decouples from the 
energy density of the mixture (\ref{3}) and can be written as
\be
\dot\rho_c+3H(\rho_c+p_c)=0.
\ee{5}
From equations (\ref{3}) and (\ref{5}) we get that the balance 
equation for the  energy density of the radiation field 
reduces to
\be
\dot\rho_r+3H(\rho_r+p_r)=-3H\varpi.
\ee{6}
In order to interpret the above equation we follow the 
work~\cite{KD1} where the  component of the 
energy-momentum pseudo-tensor of the gravitational field $T^{00}_G$ in a flat 
Robertson-Walker metric was identified with  the energy density $\rho_G$ of 
the gravitational field, i.e.,
\be
T^{00}_G\equiv\rho_G=-{3\over 8\pi G}\left({\dot a\over a}\right)^2=
-(\rho_r+\rho_c).
\ee{7}
The last equality on the right-hand side of (\ref{7})  follows from  the 
Friedmann equation (\ref{4}).
Hence, the  differentiation of (\ref{7}) thanks to (\ref{6}) leads to:
\be
\dot\rho_G+3H(\rho_G-p_r-p_c)=3H\varpi.
\ee{8}
The above equation can be interpreted as a balance equation for the 
energy density of
the gravitational field. By comparing equations (\ref{6}) and 
(\ref{8}) we conclude that the
non-equilibrium pressure  $\varpi$ is the responsible for the 
irreversible transfer of energy between the 
gravitational and radiation fields.

The relationship between the energy density of the Chaplygin gas 
and the cosmic scale factor follows from the integration
of (\ref{5}) by considering the equation of state (\ref{1}) and reads
\be
{\rho_c}=\sqrt{A+{B\over a^6}},
\ee{9}
where $B$ is a constant of integration.
 
We determine the time evolution  of the cosmic scale factor from the 
Friedmann equation 
(\ref{4}) by differentiating it with respect to time, yielding
\be
\dot H+2H^2=4\pi G\left({4A+B/a^6\over 3\sqrt{A+B/a^6}}-\varpi\right).
\ee{10}
Equation (\ref{10}) is a function of the non-equilibrium pressure and
we assume that it is a variable within the framework of extended (causal or 
second-order) thermodynamic theory. In this case the evolution equation 
for the non-equilibrium pressure -- in a linearised theory -- reads
(for a derivation of this equation from a microscopic point of view see, 
for example~\cite{CK})
\be
\varpi+\tau\dot\varpi=-3\eta H,
\ee{11}
where $\tau$ is a characteristic time while $\eta$ is the coefficient of 
bulk viscosity.

The system of equations (\ref{10}) and (\ref{11}) is closed by assuming 
that the coefficient of bulk viscosity $\eta$ and
the characteristic time $\tau$ are related to the energy densities  
by~\cite{Be,KD1}
\be
 \eta=\alpha(\rho_c+\rho_r),\qquad \tau={\eta\over (\rho_c+\rho_r)},
\ee{12}
where $\alpha$ is a constant.

In order to find the solution of the system of equations (\ref{10}) 
and (\ref{11}) it is convenient to 
write it in a dimensionless form. To this end we introduce the 
dimensionless quantities
\be
H\equiv {H\over H_0},\quad t\equiv tH_0,\quad a\equiv a
\left({A\over B}\right)^{1/6},\quad
\alpha\equiv\alpha H_0,\quad \varpi\equiv {8\pi G\varpi\over 3H_0^2},
\ee{13}
where the index zero denotes the value of the quantity at $t=0$ 
(by adjusting clocks). The Hubble parameter $H_0$ is given 
in terms of the energy densities of the Chaplygin gas and radiation at 
$t=0$, namely
\be
H_0=\sqrt{{8\pi G\over 3}(\rho_c^0+\rho_r^0)}.
\ee{14}
With respect to the above dimensionless quantities the equations 
(\ref{10}) and (\ref{11}) read
\be
\dot H+2H^2={3\over 2}\left({1\over 3(1+\rho_r^0/\rho_c^0)
\sqrt{1+1/a_0^6}}{4+1/a^6\over\sqrt{1+1/a^6}}
-\varpi\right),
\ee{15}
\be
\varpi+\alpha\dot\varpi=-3\alpha H^3.
\ee{16}

The system of differential equations  (\ref{15}) and (\ref{16}) are 
used to determine the cosmic
scale factor $a(t)$ and the non-equilibrium pressure $\varpi(t)$ once 
three initial conditions for $a(0)$, 
$\dot a(0)$ and $\varpi(0)$ are specified and values are given for the 
parameter $\alpha$ (which is connected 
with the irreversible processes) and for the ratio $\rho_r^0/\rho_c^0$ 
(which gives the order of magnitude 
between the energy densities of the Chaplygin gas and radiation at $t=0$).
From the knowledge of the time evolution of the cosmic scale factor $a(t)$ 
one can  determine the time 
evolution of the energy densities of the Chaplygin gas $\rho_c(t)$ and of 
the radiation $\rho_r(t)$, 
which follow from (\ref{9}) and  (\ref{4}), respectively,
\be
\rho_c={\sqrt{1+1/a^6}\over\sqrt{1+1/a_0^6}},\qquad
\rho_r={\rho_c^0\over\rho_r^0}\left[\left(1+{\rho_r^0\over\rho_c^0}\right)
H^2-{\sqrt{1+1/a^6}\over\sqrt{1+1/a_0^6}}\right],
\ee{17}
where $\rho_c\equiv\rho_c/\rho_c^0$ and $\rho_r\equiv\rho_r/\rho_r^0$ 
are dimensionless quantities.

If we consider the Universe as a mixture of quintessence and radiation, 
the dimensionless equations (\ref{15}) and  (\ref{17}) 
are replaced, respectively, by~\cite{KD1}
\be
\dot H+2H^2={3\over 2}\left[{(1-3w_X)\over3(1 +\rho_r^0/\rho^0_X)}
\left({1\over a}\right)^{3(w_X+1)}-\varpi\right],
\ee{18}
\be
\rho_X=\left({1\over a}\right)^{3(w_X+1)},\qquad
\rho_r=\left(1 +{\rho_X^0\over\rho^0_r}\right)H^2
-{\rho_X^0\over\rho^0_r}\left({1\over a}\right)^{3(w_X+1)},
\ee{19}
while the evolution equation for the non-equilibrium pressure (\ref{16}) 
remains unchanged. In the above 
equations $\rho_X$ is the quintessence energy density which has an equation 
of state $p_X=w_X\rho_X$. Moreover, 
according to~\cite{PR,Stein,Stein1}, the parameter $w_X$ must satisfy the 
condition $w_X<-1/3$. 

To solve the system of equations (\ref{15}) and  (\ref{16}) for the 
mixture of the Chaplygin gas and radiation, as well as the corresponding 
system (\ref{18}) and  (\ref{16}) for the mixture of quintessence and 
radiation, we specify the following initial conditions: 
$a(0)=1$ for the cosmic scale factor and  $H(0)=1$ for the Hubble 
parameter and $\dot\varpi(0)=0$ for the non-equilibrium pressure. 
There still remains much freedom to find the solutions of these two 
systems of differential equations, 
since they do depend on some parameters. For both mixtures we choose 
the parameter $\alpha=0.05$ (say). By decreasing the value of 
$\alpha$ the influence of the dissipative effects is less pronounced. 
The other parameter is the 
ratio $\rho_r^0/\rho_c^0$ (or $\rho_r^0/\rho_X^0$) which gives 
the amount of the energy density of radiation with respect to the 
energy density of the Chaplygin gas (or quintessence) 
at $t=0$. We assume that $\rho_r^0/\rho_c^0=\rho_r^0/\rho_X^0=2$ (say). 
There exists one more parameter for the case of a 
mixture of quintessence and radiation, namely $w_X$. We consider  
$w_X=-0.7$ in order to satisfy the condition $w_X<-1/3$.

\begin{figure}
\begin{center}
\includegraphics[width=7.5cm]{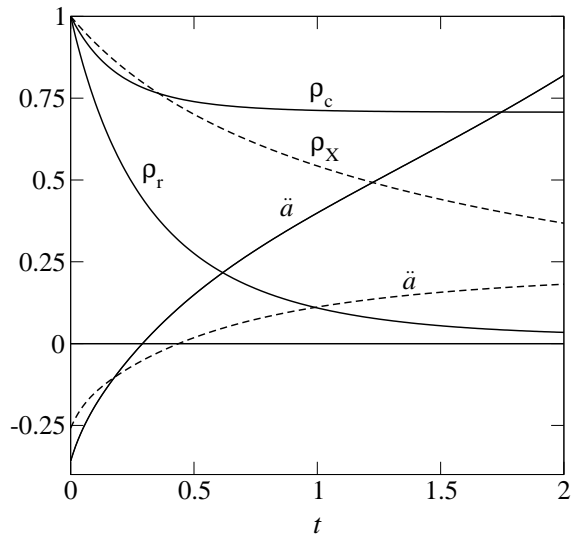}
\caption{Energy densities of Chaplygin gas $\rho_c$, of 
quintessence $\rho_X$ and   
of radiation $\rho_r$ vs time $t$.
Acceleration $\ddot a$ vs time $t$ for Chaplygin gas and radiation (straight line),
for quintessence and radiation  (dashed line).}
\end{center}
\end{figure}

In Fig. 1 we have plotted the energy density of the Chaplygin gas 
$\rho_c$, the energy density of the radiation $\rho_r$ 
and the energy density of the quintessence  $\rho_X$ as functions 
of the time $t$. Moreover, in this figure 
the acceleration $\ddot a$ is represented as a function of the time 
$t$ by a straight line 
for the mixture of a Chaplygin gas with radiation,  while  by a 
dashed line for the mixture of 
quintessence with radiation. 
From this figure we infer that for both models there exists
a period of a past deceleration with a present acceleration. 
The slope of the acceleration of the Universe modelled 
as a mixture of a Chaplygin gas with radiation is more pronounced 
than that modelled as a mixture of quintessence and 
radiation. Further, the energy density of the Chaplygin gas  tends 
to a constant value at earlier times than the energy 
density of quintessence does. The energy density of  radiation for 
both mixtures 
coincide and decays more rapidly than the energy densities of the 
Chaplygin gas and of quintessence. 
The same conclusion as in the work~\cite{KD1}  can be drawn here: 
even when a small amount of 
energy density of the Chaplygin gas (or quintessence) -- with 
respect to the  energy  density of radiation -- 
is taken into 
account, these fields evolve in such a manner that for large times the 
energy density of the Chaplygin gas (or quintessence) is very large 
with respect to the energy density 
of radiation. 
Other conclusions that can be drawn here are similar to those 
found in the work~\cite{KD1}: (a)
in the case where there is no energy density of the Chaplygin 
gas (or quintessence) only a period of deceleration is possible, 
i.e., there exists no accelerated period; (b) by decreasing  
the value of the dimensionless constant 
$\alpha$ the effect of the non-equilibrium pressure is less pronounced and
the period of past deceleration increases.

\end{document}